\documentclass[aps,pre,twocolumn,groupedaddress,showpacs]{revtex4-1}
\newcommand{\bec}[1]{\mbox{\boldmath $ #1$}}
\usepackage{graphicx}
\usepackage{bm}
\begin{document}
\title{Turbulent diffusion of chemically reacting gaseous admixtures}
\author{T. Elperin$^1$}
\email{elperin@bgu.ac.il}
\homepage{http://www.bgu.ac.il/me/staff/tov}
\author{N. Kleeorin$^{1}$}
\email{nat@bgu.ac.il}
\author{M. Liberman$^{2,3}$}
\email{misha.liberman@gmail.com}
\homepage{http://michael-liberman.com/}
\author{I. Rogachevskii$^{1}$}
\email{gary@bgu.ac.il}
\homepage{http://www.bgu.ac.il/~gary}

\medskip
\affiliation{$^1$ The Pearlstone Center for
Aeronautical Engineering Studies, Department of
Mechanical Engineering, Ben-Gurion University of
the Negev, P. O. Box 653, Beer-Sheva
84105, Israel \\
 $^2$ Nordita, KTH Royal Institute of Technology
and Stockholm University, Roslagstullsbacken 23,
10691 Stockholm, Sweden \\
 $^3$ Moscow Institute of Physics and Technology,
Dolgoprudnyi, 141700, Russia}

\date{\today}
\begin{abstract}
We study turbulent diffusion of chemically
reacting gaseous admixtures in a developed
turbulence. In our previous study [Phys. Rev.
Lett. {\bf 80}, 69 (1998)] using a path-integral
approach for a delta-correlated in time random
velocity field, we demonstrated a strong
modification of turbulent transport in fluid
flows with chemical reactions or phase
transitions. In the present study we use the
spectral tau approximation, that is valid for
large Reynolds and Peclet numbers, and show that
turbulent diffusion of the reacting species can
be strongly depleted by a large factor that is
the ratio of turbulent and chemical times (turbulent Damk\"{o}hler number).
We have demonstrated that the derived theoretical dependence of turbulent diffusion
coefficient versus the turbulent Damk\"{o}hler number
is in a good agreement with that obtained previously in the numerical modelling
of a reactive front propagating in a turbulent flow and
described by the Kolmogorov-Petrovskii-Piskunov-Fisher equation.
We have found that turbulent
cross-effects, e.g., turbulent mutual diffusion
of gaseous admixtures and turbulent Dufour-effect
of the chemically reacting gaseous admixtures,
are less sensitive to the values of
stoichiometric coefficients. The mechanisms of
the turbulent cross-effects are different from
the molecular cross effects known in irreversible
thermodynamics. In a fully developed turbulence
and at large Peclet numbers the turbulent
cross-effects are much larger than the molecular
ones. The obtained results are applicable also to
heterogeneous phase transitions.
\end{abstract}


\maketitle

\section{Introduction}

Turbulent transport in flows with chemical
reactions is of great interest in various
applications, ranging from combustion to physics
of turbulent atmosphere of the Earth (see, e.g.,
\cite{ZB85,F03,P04,EM11,SB11,SP06,PK97,W99,J05}). During the
decades turbulent transport of passive scalar and
particles has been subject of an active research
(see, e.g., handbooks
\cite{MY75,CSA80,Mc90,ZRS90,BLA97,CST11} and
reviews \cite{WA00,S03,BH03,KPE07,WA09,BE10}).
Many important problems, including particle
clustering in isothermal
\cite{EKR96a,BF01,EKR02,FP04,EKR07,SA08,XB08,SSA08}
and stratified \cite{EKR10,EKR13} turbulence,
intermittency \cite{KR68,F95}, effective diffusion \cite{DF14},
the formation of large-scale inhomogeneous structures in spatial
distribution of particles or different scalar
fields in small-scale turbulence
\cite{EKR96,EKR00,PM02,BEE04,RE05,EEKR06,SSEKR09,HKRB12}
have been investigated in analytical, numerical
and laboratory studies. However, impact of
chemical reactions on  turbulent transport have
been studied mainly numerically and in the
context of turbulent combustion (see, e.g.,
\cite{F03,P04,EM11,SB11}).

Combustion process is the chemical reaction accompanied
by heat release. Turbulent combustion
can proceed as volume distributed chemical reaction
(e.g., as a homogeneous burning of the turbulent
premixed gaseous mixture) or propagate as
a flame front in a turbulent flow separating fresh
unburned fuel and combustion products (see, e.g., \cite{D08}). In both
cases turbulence is created by an external forcing
and can be enhanced by intrinsic instability of the
flame front (see, e.g., \cite{ZB85,P04,VV02,HT04}).

Notice that temperature and equilibrium composition of
combustion products are purely thermodynamic characteristics
determined by the thermodynamic equilibrium laws which yield a
relation between the initial and final stages. In this case
chemical kinetics described by a one-step Arrhenius model
provides results which usually are in a good agreement with
experimental data. On the contrary, in order to reproduce
transient processes that are accompanied by compression and
shock waves, it is necessary to take into account the detailed
chemical reaction mechanisms involving several hundreds chemical reactions.
This complex chemistry determines the chemical time scales,
such as induction time and period of exothermal reaction, competing with
transport time scales in formation of the zone of energy release.
The latter determines the evolution of the propagating flame (see, e.g., \cite{ML14}).

In a turbulent atmosphere the most common are the volume
distributed chemical reactions, while
during wild fires propagation of the turbulent
flame front is of particular interest in the atmospheric
and industrial applications (see, e.g., \cite{SP06,PK97,W99,J05}).
In the previous studies the main
attention was focused on evaluation of nonlinear
sources in the governing equations for concentrations
of chemical species. In such description the
effect of chemical reactions on turbulent transport
coefficients has been neglected. For the
first time the effect of chemistry on turbulent diffusion
was studied in \cite{EKR98}
by means of a path-integral approach for the
Kraichnan-Kazantsev model (see \cite{KR68,K68})
of the random velocity field, and it was found that
turbulent diffusion can be strongly depleted by
chemical reactions or phase transitions. It was
also shown in \cite{EKR98} that there exists an
additional non-diffusive turbulent flux of number
density of gaseous admixture (proportional to the
mean temperature gradient multiplied by the
number density of gaseous admixture) and
additional turbulent heat flux (proportional to
the gradient of the mean number density of
gaseous admixture) in flows with chemical
reactions or phase transitions.

The effect of chemistry on turbulent diffusion
has been recently studied in \cite{BH11} using mean-field
simulations (MFS) and direct numerical simulations
(DNS). In these simulations a reactive
front propagation in a turbulent flow was investigated
using the Kolmogorov-Petrovskii- Piskunov
(KPP) equation \cite{KPP37} or the Fisher equation \cite{F37}.
This equation has also been amended by an advection
term to describe the interaction with a
turbulent velocity field \cite{BN09,PBN10}. In MFS of \cite{BH11}
memory effects of turbulent diffusion have been
taken into account to determine the front speed
in the case when the turbulent time, $\tau_0$, is much
larger than the characteristic chemical time, $\tau_c$. It
was found that the memory effects saturate the
front speed to values of the turbulent speed, while
the nonlinearity of the reaction term increases the
front speed. This study allows to determine the dependence of the
turbulent diffusion coefficient versus the turbulent
Damk\"{o}hler number, ${\rm Da}_{_{\rm
T}}=\tau_0/\tau_c$.

In the present study we investigate turbulent
transport of chemically reacting gaseous
admixtures in a developed turbulence using a
spectral tau approach (high-order closure
procedure), see, e.g., \cite{O70,PFL76,Mc90}. We
have demonstrated here the existence of the
turbulent cross-effects, including turbulent
mutual diffusion of gaseous admixtures and
turbulent Dufour-effect. The mechanisms of these
cross-effects are different from the molecular
cross effects known in irreversible
thermodynamics (see, e.g., \cite{GM84}). In a
developed turbulence and at large Peclet numbers
the turbulent cross-effects are much larger than
the molecular ones.
These results are also valid for heterogeneous
phase transitions. We show that for a large
turbulent Damk\"{o}hler number ${\rm Da}_{_{\rm
T}} \gg 1$ turbulent diffusion of
the admixtures can be strongly reduced by a large
factor ${\rm Da}_{_{\rm T}}$ depending on the
value of stoichiometric coefficients of chemical
species. In this paper we illustrate effect
of chemistry on turbulent diffusion using simple
global one step chemical reactions taking into account
the reaction order. Such approach provides qualitatively
and often quantitatively correct physics.

The paper is organized as follows. The governing
equations and applied methods are formulated in
Section II. Turbulent fluxes of chemically
reacting admixtures are determined in Sections
III and IV. Comparison of theoretical predictions
with numerical simulations
is given in Section V. Finally, in Section VI we draw
conclusions and discuss the implications of the
obtained results.

\section{Governing equations}

Advection-diffusion equation for the number
density $n_\beta(t,{\bm x})$ of chemically
reactive admixtures in a turbulent flow reads:
\begin{eqnarray}
\frac{\partial n_\beta}{\partial t} + {\bm \nabla
\cdot} (n_\beta \,{\bm v}) = - \nu_\beta \hat
W(n_\beta, T) + \hat D(n_\beta),
 \label{B1}
\end{eqnarray}
where ${\bm v}(t,{\bm x})$ is the instantaneous
fluid velocity field, $\hat D(n_\beta) =$
div $[\rho \, D_\beta {\bm \nabla} (n_\beta / \rho)]$
is the linear diffusion operator of $n_\beta$
(see, e. g.,  \cite{LL87}), $D_\beta$ is the coefficient
of the molecular diffusion based on molecular Fick's law, $T$ is the fluid temperature,
$- \nu_\beta \hat W(n_\beta, T)$ is the source (or sink) term and $\nu_\beta$
is the stoichiometric coefficient that is the order of the
reaction with respect to species $\beta$, $\sum_{\beta=1}^m \, \nu_\beta$
is the overall order of the reaction
and $m$ is total number of species.
The function $\hat W(n_\beta, T)$ satisfies to the Arrhenius law
(see, e.g., \cite{ML08}):
\begin{eqnarray}
\hat W=A \exp \left(- E_a/R \, T \right) \,
\Large {\Pi}_{\beta=1}^m \, (n_\beta)^{\nu_\beta} ,
 \label{BB1}
\end{eqnarray}
where $A$ is the reaction rate constant, $E_a$ is
the activation energy and $R$ is the universal
gas constant. Evolution of the temperature field
$T(t,{\bm x})$ in a turbulent fluid flow is
determined by the following equation
\begin{eqnarray}
{\partial T \over \partial t} &+& ({\bm v} \cdot
{\bm  \nabla}) T + (\gamma - 1) T ({\bm \nabla}
\cdot {\bm v})
 \nonumber\\
&&= q \hat W(n_\beta, T) + \hat D(T),
 \label{B2}
\end{eqnarray}
where the term $\hat D(T)= \rho^{-1}$ \, div $[\rho \, \chi {\bm \nabla} T]$
determines the molecular diffusion of the fluid temperature,
$\chi$ is the coefficient of molecular diffusion of temperature,
$q=Q/ \rho c_p$, $Q$ is the reaction energy
release, $c_p$ is the specific heat at constant
pressure, $\rho$ is the fluid density and
$\gamma=c_p/c_v$ is the ratio of specific heats.
The density $\rho$ and the velocity ${\bm v}$ of
the fluid satisfy the continuity equation
\begin{eqnarray}
{\partial \rho \over \partial t} + {\bm
\nabla}\cdot(\rho {\bm v}) = 0 ,
 \label{B3}
\end{eqnarray}
and the Navier-Stokes equation:
\begin{eqnarray}
{\partial {\bm v} \over \partial t} &+& ({\bm v}
\cdot {\bm  \nabla}) {\bm v} = - {1 \over \rho}
{\bm  \nabla} P + \hat D_\nu({\bm v}),
 \label{BB3}
\end{eqnarray}
and $\hat D_\nu({\bm v})$ is the viscous term and
$P$ is the fluid pressure.

To study the formation of large-scale
inhomogeneous structures, Eqs.~(\ref{B1})
and~(\ref{B2}) are averaged over an ensemble of
turbulent velocity fields. Using a mean-field
approach, we decompose $n_\beta$ and $T$ into the
mean quantities, $\overline{N}_\beta$ and
$\overline{T}$, and fluctuations $n'_\beta$ and
$\theta$, where $\overline{n'_\beta}=0$ and
$\overline{\theta}=0$. We decompose the velocity
field in a similar fashion, and assume for
simplicity vanishing mean fluid velocity,
$\overline{\bm U}=0$. Averaging Eqs.~(\ref{B1})
and~(\ref{B2}) over an ensemble of turbulent
velocity fields we obtain the equations for the
mean fields:
\begin{eqnarray}
\frac{\partial \overline{N}_\beta}{\partial t}
&+& {\bm \nabla \cdot} \langle n'_\beta \,{\bm u}
\rangle = - \nu_\beta \overline{W} + \hat
D(\overline{N}_\beta),
 \label{EE3}\\
\frac{\partial \overline{T}}{\partial t} &+& {\bm
\nabla \cdot} \langle \theta \,{\bm u} \rangle +
(\gamma - 2) \langle \theta ({\bm \nabla} \cdot
{\bm u}) \rangle
 \nonumber\\
&&= q \overline{W} + \hat D(\overline{T}) ,
 \label{EE4}
\end{eqnarray}
where ${\bm u}$ are the fluid velocity
fluctuations, $\overline{W} \equiv \langle \hat
W(n_\beta, T) \rangle$ and angular brackets imply
the averaging over the statistics of turbulent
velocity field. To obtain a closed system of the
mean-field equations one needs to determine the
turbulent fluxes $\langle n'_\beta \,{\bm u}
\rangle$ and $\langle \theta \,{\bm u} \rangle$
as well as $\langle \theta ({\bm \nabla} \cdot
{\bm u}) \rangle$. To this end we use the
following equations for fluctuations
$n'_\beta=n_\beta-\overline{N}_\beta$ and
$\theta=T-\overline{T}$:
\begin{eqnarray}
&&\frac{\partial n'_\beta}{\partial t} + {\bm
\nabla \cdot} \Big(n'_\beta \,{\bm u}-\langle
n'_\beta \,{\bm u} \rangle\Big) = - \nu_\beta
\Big(\hat W -\overline{W} \Big)
\nonumber\\
&&\quad - {\bm \nabla \cdot} (\overline{N}_\beta
\,{\bm u}) + \hat D(n'_\beta),
 \label{EE5}\\
&&\frac{\partial \theta}{\partial t} + {\bm
\nabla \cdot} \Big(\theta \,{\bm u}-\langle
\theta \,{\bm u} \rangle\Big) + (\gamma - 2)
\Big[\theta ({\bm \nabla} \cdot {\bm u}) -
\langle \theta ({\bm \nabla} \cdot {\bm u})
\rangle \Big]
 \nonumber\\
&&\quad = q \left[\hat W - \overline{W} \right] -
({\bm u} \cdot {\bm \nabla}) \overline{T} -
(\gamma - 1) \overline{T} ({\bm \nabla} \cdot
{\bm u}) + \hat D(\theta) .
 \nonumber\\
 \label{EE6}
\end{eqnarray}
These equations are obtained by subtracting
Eqs.~(\ref{EE3}) and~(\ref{EE4}) for the mean
fields from Eqs.~(\ref{B1}) and~(\ref{B2}) for
the total fields.

Now we assume that the temperature fluctuations
are much smaller than the mean temperature, and the
fluctuations of the number density of admixtures
are much smaller than mean values. This allows us to
expand the source term in a series:
\begin{eqnarray}
\hat W - \overline{W} &=& \sum_{\beta=1}^{m}
\left({\partial \hat W \over
\partial n_\beta} \right)_{\overline{N}_\beta}
n'_\beta + \left({\partial \hat
W \over \partial T}\right)_{\overline{T}} \theta
 \nonumber\\
&+& O\left[(n'_\beta)^{2}; \theta^2; n'_\beta \,
\theta\right] \equiv \overline{W} (C'_n + C'_T) ,
 \label{B4}
\end{eqnarray}
where we take into account that for the Arheneus
law~(\ref{BB1}):
\begin{eqnarray}
\left({\partial \ln \hat W \over \partial
n_\beta} \right)_{\overline{N}_\beta}= {\nu_\beta
\over \overline{N}_\beta} ,
 \quad
\left({\partial \ln \hat W \over \partial
T}\right)_{\overline{T}} = {E_a \over R
\overline{T}^2} .
 \label{B5}
\end{eqnarray}
Here we introduced the following new variables:
\begin{eqnarray}
C'_n &=& \sum_{\beta=1}^{m} {\nu_\beta \over
\overline{N}_\beta} n'_\beta ,
 \quad
C'_T = {E_a \over R \overline{T}^2} \theta ,
 \label{B6}
\end{eqnarray}
where the evolutionary equations for $C'_n$ and
$C'_T$ read:
\begin{eqnarray}
&&\frac{\partial C'_n}{\partial t} + {\bm \nabla
\cdot} \Big(C'_n \,{\bm u}-\langle C'_n \,{\bm u}
\rangle\Big) = - \alpha_n \overline{W} (C'_n +
C'_T)
\nonumber\\
&&\quad - \sum_{\beta=1}^{m} {\nu_\beta \over
\overline{N}_\beta} {\bm \nabla \cdot}
(\overline{N}_\beta \,{\bm u}) + \hat D(C'_n),
 \label{C3}\\
&&\frac{\partial C'_T}{\partial t} + {\bm \nabla
\cdot} \Big(C'_T \,{\bm u}-\langle C'_T \,{\bm u}
\rangle\Big) + (\gamma - 2) \Big[C'_T ({\bm
\nabla} \cdot {\bm u})
 \nonumber\\
&&\quad - \langle C'_T ({\bm \nabla} \cdot {\bm
u}) \rangle \Big] = \alpha_{_{T}} \overline{W}
(C'_n + C'_T) + \hat D(C'_T)
 \nonumber\\
&&\quad - {\alpha_{_{T}}\over q}\, \left[({\bm u}
\cdot {\bm \nabla}) \overline{T} + (\gamma - 1)
\overline{T} ({\bm \nabla} \cdot {\bm u}) \right]
,
 \label{C4}
\end{eqnarray}
where
\begin{eqnarray}
\alpha_n &=& \sum_{\beta=1}^{m} {\nu_\beta \over
\overline{N}_\beta} ,
 \quad
\alpha_{_{T}} = {q E_a \over R \overline{T}^2}.
 \label{B9}
\end{eqnarray}

Using Eqs.~(\ref{C3})-(\ref{C4}) and the
Navier-Stokes equation~(\ref{BB3}) we derive
equations for the second-order moments $\langle
C'_n \,{\bm u} \rangle$ and $\langle C'_T \,{\bm
u} \rangle$:
\begin{eqnarray}
&&\frac{\partial \langle C'_n \,u_i
\rangle}{\partial t} = - \alpha_n \overline{W}
\Big(\langle C'_n \,u_i \rangle + \langle C'_T
\,u_i \rangle\Big) + \hat {\cal N} \langle C'_n
\,u_i \rangle
\nonumber\\
&&\quad - \sum_{\beta=1}^{m} \nu_\beta \Big[
\langle u_i \,u_j \rangle \, \nabla_j \ln
\overline{N}_\beta + \langle u_i({\bm \nabla}
\cdot {\bm u}) \rangle \Big] ,
 \label{C5}\\
&&\frac{\partial \langle C'_T \,u_i
\rangle}{\partial t} = {\alpha_{_{T}}\over q}  \,
\Big[q\overline{W} \left(\langle C'_n \,u_i
\rangle + \langle C'_T \,u_i \rangle\right) -
\langle u_i \,u_j \rangle \, \nabla_j
\overline{T}
\nonumber\\
&&\quad  - (\gamma - 1) \overline{T} \, \langle
u_i({\bm \nabla} \cdot {\bm u}) \rangle \Big] +
\hat {\cal N} \langle C'_T \,u_i \rangle,
 \label{C6}
\end{eqnarray}
where $\hat {\cal N} \langle C'_n \,u_i \rangle$
and $\hat {\cal N} \langle C'_T \,u_i \rangle$
include the third-order moments caused by the
nonlinear terms and the second-order moments due
to the molecular dissipative terms:
\begin{eqnarray}
&& \hat {\cal N} \langle C'_n \,u_i \rangle = -
\langle [{\bm \nabla \cdot} (C'_n \,{\bm u})]
\,u_i \rangle + \langle \hat D(C'_n) \,u_i
\rangle
\nonumber\\
&&\quad - \langle C'_n \, [({\bm u}\cdot {\bm
\nabla}) u_i + \rho^{-1} \nabla_i p'] \rangle  +
\langle C'_n \, \hat D_\nu(u_i) \rangle ,
 \label{CC5}\\
&& \hat {\cal N} \langle C'_T \,u_i \rangle = -
\langle [{\bm \nabla \cdot} (C'_T \,{\bm u}) +
(\gamma - 2) C'_T ({\bm \nabla} \cdot {\bm u})]
\,u_i \rangle
\nonumber\\
&&\quad + \langle \hat D(C'_T) \,u_i \rangle -
\langle C'_T \, [({\bm u}\cdot {\bm \nabla}) u_i
+ \rho^{-1} \nabla_i p'] \rangle
\nonumber\\
&&\quad + \langle C'_n \, \hat D_\nu(u_i) \rangle
,
 \label{CC6}
\end{eqnarray}
and $p'$ are the fluid pressure fluctuations.

In the anelastic approximation for the low-
Mach-number flows (Ma $ = u/c_s \ll 1)$ and for
hydrodynamic times that are much larger than
the acoustic times, the continuity equation for surrounding fluid reads
${\bm \nabla} \cdot {\bm u}= -({\bm u}\cdot
{\bm \nabla}) \ln \rho$. We consider the ideal
gases. Therefore,
\begin{eqnarray}
\overline{T} \, \langle u_i({\bm \nabla} \cdot
{\bm u}) \rangle = \langle u_i \,u_j \rangle \,
(\nabla_j \ln \overline{P} -\nabla_j \ln \overline{P}).
 \label{CC7}
\end{eqnarray}
In a particular case when there is no external
pressure gradient, ${\bm \nabla} \overline{P}=0$, (e.g.,
there is no mean flow), Eq.~(\ref{CC7})
yields:
\begin{eqnarray}
\overline{T} \, \langle u_i({\bm \nabla} \cdot
{\bm u}) \rangle = \langle u_i \,u_j \rangle \,
\nabla_j \overline{T},
 \label{C7}
\end{eqnarray}
where $\overline{P}$ is the mean fluid pressure.
This allows us to rewrite
Eqs.~(\ref{C5})-(\ref{C6}) in the following form:
\begin{eqnarray}
&&\frac{\partial \langle C'_n \,u_i
\rangle}{\partial t} = - \alpha_n \overline{W}
\Big(\langle C'_n \,u_i \rangle + \langle C'_T
\,u_i \rangle\Big) + \hat {\cal N} \langle C'_n
\,u_i \rangle
\nonumber\\
&&\quad - \langle u_i \,u_j \rangle \, \nabla_j
\sum_{\beta=1}^{m} \nu_\beta \Big[\ln
\overline{N}_\beta + \ln \overline{T} \Big] ,
 \label{C10}\\
&&\frac{\partial \langle C'_T \,u_i
\rangle}{\partial t} = \alpha_{_{T}}
\Big[\overline{W} \left(\langle C'_n \,u_i
\rangle + \langle C'_T \,u_i \rangle\right) -
{\gamma\over q}  \, \langle u_i \,u_j \rangle \,
\nabla_j \overline{T} \Big]
\nonumber\\
&&\quad + \hat {\cal N} \langle C'_T \,u_i
\rangle .
 \label{C8}
\end{eqnarray}
Now we introduce a new variable $C'=C'_n+C'_T$.
Equations~(\ref{C3})-(\ref{C4}) allow us to
derive equation for the second-order moment
$\langle C' \,u_i \rangle$:
\begin{eqnarray}
&&\frac{\partial \langle C' \,u_i
\rangle}{\partial t} = - \tau_c^{-1} \, \langle
C' \,u_i \rangle + \hat {\cal N} \langle C' \,u_i
\rangle - \langle u_i \,u_j \rangle
\nonumber\\
&&\, \times \Big[\sum_{\beta=1}^{m} \nu_\beta \,
\nabla_j \Big(\ln \overline{N}_\beta + \ln
\overline{T} \Big) + {\gamma\over q} \,
\alpha_{_{T}} \, \nabla_j\overline{T} \Big],
 \label{C9}
\end{eqnarray}
where $\tau_c^{-1}=\overline{W} (\alpha_n
-\alpha_{_{T}})$ is the inverse chemical time,
$\hat {\cal N} \langle C' \,u_i \rangle=\hat
{\cal N} \langle C'_n \,u_i \rangle+\hat {\cal N}
\langle C'_T \,u_i \rangle$ and the third-order
moments $\hat {\cal N} \langle C'_n \,u_i
\rangle$ and $\hat {\cal N} \langle C'_T \,u_i
\rangle$ are determined by
Eqs.~(\ref{CC5})-(\ref{CC6}).

It should be noted that the anelastic approximation is
used in the continuity equation for the surrounding fluid. The chemical
time appears only in the equation for the number
density for species and the temperature equation. Since
the ratio of spatial density of species is assumed to be
much smaller than the density of the surrounding fluid (i.e., small mass-loading
parameter), there is only one-way coupling, i.e.,
no effect of species on the
fluid flow. Due to the same reason the energy release
(or absorbtion of energy) caused by chemical reactions is
much smaller than the internal energy of the surrounding
fluid. This implies that even small chemical time does
not affect the fluid characteristics.

\section{Turbulent flux of $C$}

The equation for the second-order
moment~(\ref{C9}) includes the first-order
spatial differential operators applied to the
third-order moments $\hat{\cal N} F^{(III)}$ [see
Eqs.~(\ref{CC5})-(\ref{CC6})]. To close the
system of equations it is necessary to express
the third-order terms $\hat{\cal N} F^{(III)}$
through the lower-order moments $F^{(II)}$ (see,
e.g., \cite{O70,MY75,Mc90}). We use the spectral
$\tau$ approximation which postulates that the
deviations of the third-order moments, $\hat{\cal
N} F^{(III)}({\bm k})$, from the contributions to
these terms afforded by the background
turbulence, $\hat{\cal N} F^{(III,0)}({\bm k})$,
can be expressed through the similar deviations
of the second-order moments, $F^{(II)}({\bm k}) -
F^{(II,0)}({\bm k})$:
\begin{eqnarray}
&& \hat{\cal N} F^{(III)}({\bm k}) - \hat{\cal N}
F^{(III,0)}({\bm k})
\nonumber\\
&& \quad \quad\quad = - {1 \over \tau_r(k)} \,
\Big[F^{(II)}({\bm k}) - F^{(II,0)}({\bm
k})\Big],
 \label{D1}
\end{eqnarray}
(see, e.g., \cite{O70,PFL76,RK07}), where
$\tau_r(k)$ is the scale-dependent relaxation
time, which can be identified with the
correlation time $\tau(k)$ of the turbulent
velocity field for large Reynolds and Peclet
numbers. The functions with the superscript $(0)$
correspond to the background turbulence with zero
gradients of the mean temperature, the mean
number density and the mean fluid density.

In order to elucidate the spectral $\tau$
approach in the following we present the explicit
form of terms in Eq.~(\ref{D1}) corresponding to
the second-order moments, $F^{(II)}({\bf
k})\equiv\langle C'_n({\bm k}) \,u_i(-{\bm k})
\rangle$, and the third-order moments $\hat{\cal
N} F^{(III)}({\bm k}) \equiv \hat{\cal N} \langle
C'_n \,u_i\rangle_{\bm k}$:
\begin{eqnarray}
&& \hat{\cal N} \langle C'_n \,u_i\rangle_{\bm k}
- \hat{\cal N} \langle C'_n \,u_i\rangle_{\bm
k}^{(0)}
\nonumber\\
&& \quad = - {1 \over \tau(k)} \, \Big[\langle
C'_n({\bm k}) \,u_i(-{\bm k}) \rangle - \langle
C'_n({\bm k}) \,u_i(-{\bm k}) \rangle^{(0)}
\Big].
\nonumber\\
 \label{DD1}
\end{eqnarray}
Similarly one can formulate Eq.~(\ref{D1}) for
all other third-order and second-order
correlators involving velocity, temperature and
species number density fluctuations in ${\bf
k}$-space. Validation of the $\tau$ approximation
for different situations has been performed in
numerous numerical simulations and analytical
studies (see, e.g., review \cite{BS05}; and also
discussion in \cite{RK07}, Sec. 6).

The $\tau$-approximation is in general similar to
Eddy Damped Quasi Normal Markowian (EDQNM)
approximation. However some principle difference
exists between these two approaches (see
\cite{O70,Mc90}). The EDQNM closures do not relax
to equilibrium, and this procedure does not
describe properly the motions in the equilibrium
state in contrast to the $\tau$-approximation.
Within the EDQNM theory, there is no dynamically
determined relaxation time, and no slightly
perturbed steady state can be approached
\cite{O70}. In the $\tau$-approximation, the
relaxation time for small departures from
equilibrium is determined by the random motions
in the  equilibrium state, but not by the
departure from equilibrium. We use the
$\tau$-approximation, but not the EDQNM
approximation because we consider a case when the
characteristic scales of variations of the mean
number density of admixtures and the mean
temperature are much larger than the integral
turbulence scale. Analysis performed in
\cite{O70} showed that the $\tau$-approximation describes the
relaxation to equilibrium state significantly more
accurately than the EDQNM approach.

The $\tau$ approach is an universal tool in
turbulent transport that allows to obtain closed
results and compare them with the results of
laboratory experiments, observations and
numerical simulations. The $\tau$ approximation
reproduces many well-known phenomena found by
other methods in turbulent transport of particles
and magnetic fields, in turbulent convection and
stably stratified turbulent flows
\cite{BS05,RK07,RKKB11}.

Note that when the gradients of the mean
temperature and the mean number density are zero,
the turbulent heat flux and the turbulent flux of
chemical admixtures vanish, and the contributions
of the corresponding fluctuations [the terms with
the superscript (0)], vanish as well.
Consequently, Eq.~(\ref{DD1}) reduces to
$\hat{\cal N} \langle C'_n \,u_i\rangle_{\bm k} =
- \langle C'_n({\bm k}) \,u_i(-{\bm k}) \rangle /
\tau(k)$. We also assume that the characteristic
time of variation of the second-order moments are
substantially larger than the correlation time
$\tau(k)$ for all turbulence scales. Therefore,
the steady-state version of Eq.~(\ref{C9}),
written in the Fourier space yields the following
formulae for the turbulent flux $\langle C'({\bm
k}) \,u_i(-{\bm k}) \rangle$:
\begin{eqnarray}
&& \langle C'({\bm k}) \,u_i(-{\bm k}) \rangle =
- \tau_{\rm eff}(k) \, \langle u_i({\bm k}) \,
u_j(-{\bm k}) \rangle \,
\nonumber\\
&&\quad \quad \times \Big[\sum_{\beta=1}^{m}
\nu_\beta \, \nabla_j \Big(\ln \overline{N}_\beta
+ \ln \overline{T} \Big) + {\gamma\over q}  \,
\alpha_{_{T}} \, \nabla_j\overline{T} \Big],
 \label{D2}
\end{eqnarray}
where $\tau^{-1}_{\rm eff}(k) =
\tau^{-1}_{c}+\tau^{-1}(k)$.

To integrate in ${\bm k}$-space we need to choose a model
of the background turbulence.
In order to separate the turbulent transport effects caused by the chemistry
from those caused by inhomogeneity of turbulence we consider isotropic
and homogeneous background turbulence, $\langle u_i({\bm k}) \, u_j(-{\bm
k}) \rangle$ (see, e.g., \cite{B53}):
\begin{eqnarray}
\langle u_i({\bm k}) \, u_j(-{\bm k}) \rangle =
{u_0^2 \, E_T(k) \over 8 \pi k^2}
\Big[\delta_{ij} - {k_i \, k_j \over k^2} \Big],
 \label{D3}
\end{eqnarray}
where $E_T(k)=(\mu-1) k_{0}^{-1} (k /
k_{0})^{-\mu}$ is the energy spectrum function
with the exponent $1<\mu<3$, $\tau(k) = 2 \,
\tau_0 \, (k / k_{0})^{1-\mu}$ is the turbulent
correlation time, $\tau_0 = \ell / u_{0}$ is the
characteristic turbulent time and $u_{0}$ is the
characteristic turbulent velocity in the integral
scale $\ell$.

After integration in ${\bm k}$ space we obtain
the turbulent flux $\langle C'\,u_i \rangle =
\int \langle C'({\bm k}) \,u_i(-{\bm k}) \rangle
\, d {\bm k}$:
\begin{eqnarray}
\langle C'\,u_i \rangle &=& - D^T_C\,
\Big[\sum_{\beta=1}^{m} \nu_\beta \, \nabla_j
\Big(\ln \overline{N}_\beta + \ln \overline{T}
\Big)
\nonumber\\
&&+ {\gamma\over q} \,\alpha_{_{T}}\,
 \nabla_j\overline{T} \Big],
 \label{D4}
\end{eqnarray}
where $D^T_0 = \tau_0 u_0^2/3$, and the turbulent
diffusion coefficient reads:
\begin{eqnarray}
D^T_C=D^T_0 \, \left(1- {\ln(1 + 2{\rm Da}_{_{\rm
T}}) \over 2{\rm Da}_{_{\rm T}}} \right) ,
 \label{D5}
\end{eqnarray}
where ${\rm Da}_{_{\rm T}}=\tau_0/\tau_c$ is the
turbulent Damk\"{o}hler number. Here we used that
$\int \tau_{\rm eff}(k) \langle u_i({\bm k}) \,
u_j(-{\bm k}) \rangle \, d{\bm k} = D^T_C
\delta_{ij}$. The asymptotic behaviour of the
turbulent diffusion coefficient is as follows:
when ${\rm Da}_{_{\rm T}} \ll 1$, the turbulent
diffusion coefficient is
\begin{eqnarray}
D^T_C=D^T_0 \, {\rm Da}_{_{\rm T}} \left(1-
{4{\rm Da}_{_{\rm T}} \over 3} \right),
 \label{D6}
\end{eqnarray}
while for ${\rm Da}_{_{\rm T}} \gg 1$, it is
\begin{eqnarray}
D^T_C=D^T_0 \, \left(1- {\ln{\rm Da}_{_{\rm
T}}\over 2 {\rm Da}_{_{\rm T}}} \right) .
 \label{D7}
\end{eqnarray}
Equations~(\ref{D2}) and~(\ref{D4})-(\ref{D7})
allow us to determine the turbulent transport
coefficients for gaseous admixtures and
temperature field (see next section).

\section{Turbulent transport of
admixtures and temperature}

Using Eqs.~(\ref{B4})-(\ref{B5}) we rewrite
Eqs.~(\ref{EE5})-(\ref{EE6}) for the fluctuations
of the number density of admixtures,
$n'_\beta=n_\beta - \overline{N}_\beta$, and the
fluid temperature field, $\theta=T-\overline{T}$,
in the following form:
\begin{eqnarray}
&&\frac{\partial n'_\beta}{\partial t} + {\bm
\nabla \cdot} \Big(n'_\beta \,{\bm u}-\langle
n'_\beta \,{\bm u} \rangle\Big) = - \nu_\beta
\overline{W} (C'_n + C'_T)
\nonumber\\
&&\quad - {\bm \nabla \cdot} (\overline{N}_\beta
\,{\bm u}) + \hat D(n'_\beta),
 \label{E5}\\
&&\frac{\partial \theta}{\partial t} + {\bm
\nabla \cdot} \Big(\theta \,{\bm u}-\langle
\theta \,{\bm u} \rangle\Big) + (\gamma - 2)
\Big[\theta ({\bm \nabla} \cdot {\bm u}) -
\langle \theta ({\bm \nabla} \cdot {\bm u})
\rangle \Big]
 \nonumber\\
&&\quad  = q \overline{W} (C'_n + C'_T) - ({\bm
u} \cdot {\bm \nabla}) \overline{T} - (\gamma -
1) \overline{T} ({\bm \nabla} \cdot {\bm u})
 \nonumber\\
&&\quad + \hat D(\theta) .
 \label{E6}
\end{eqnarray}
To close a system of the mean-field equations we
determine the turbulent fluxes $\langle n'_\beta
\,{\bm u} \rangle$ and $\langle \theta \,{\bm u}
\rangle$. Equations for these second moments
read:
\begin{eqnarray}
\frac{\partial \langle n'_\beta \,u_i
\rangle}{\partial t} &=& - \nu_\beta \overline{W}
\langle C' \,u_i \rangle - \langle u_i \,u_j
\rangle \, \nabla_j \overline{N}_\beta
\nonumber\\
&&  - \overline{N}_\beta \, \langle u_i({\bm
\nabla} \cdot {\bm u}) \rangle + \hat {\cal N}
\langle n'_\beta \,u_i \rangle,
 \label{E7}\\
\frac{\partial \langle \theta \,u_i
\rangle}{\partial t} &=& q \overline{W} \langle
C' \,u_i \rangle - \langle u_i \,u_j \rangle \,
\nabla_j \overline{T}
\nonumber\\
&&  - (\gamma - 1) \overline{T} \, \langle
u_i({\bm \nabla} \cdot {\bm u}) \rangle + \hat
{\cal N} \langle \theta \,u_i \rangle .
 \label{E8}
\end{eqnarray}
where $\hat {\cal N} \langle n'_\beta \,u_i
\rangle$ and $\hat {\cal N} \langle \theta \,u_i
\rangle$ include the third-order moments caused
by the nonlinear terms and the second-order
moments due to the dissipative terms in
Eqs.~(\ref{E5})-(\ref{E6}) and the Navier-Stokes
equation:
\begin{eqnarray}
&& \hat {\cal N} \langle n'_\beta \,u_i \rangle =
- \langle [{\bm \nabla \cdot} (n'_\beta \,{\bm
u})] \,u_i \rangle + \langle \hat D(n'_\beta)
\,u_i \rangle
\nonumber\\
&&\quad - \langle n'_\beta \, [({\bm u}\cdot {\bm
\nabla}) u_i + \rho^{-1} \nabla_i p'] \rangle  +
\langle n'_\beta \, \hat D_\nu(u_i) \rangle ,
 \label{EC5}\\
&& \hat {\cal N} \langle \theta \,u_i \rangle = -
\langle [{\bm \nabla \cdot} (\theta \,{\bm u}) +
(\gamma - 2) \theta ({\bm \nabla} \cdot {\bm u})]
\,u_i \rangle
\nonumber\\
&&\quad + \langle \hat D(\theta) \,u_i \rangle -
\langle \theta \, [({\bm u}\cdot {\bm \nabla})
u_i + \rho^{-1} \nabla_i p'] \rangle
\nonumber\\
&&\quad + \langle \theta \, \hat D_\nu(u_i)
\rangle .
 \label{EC6}
\end{eqnarray}

Next, we apply the $\tau$ approximation to
Eqs.~(\ref{E7})-(\ref{E8}), written in ${\bf
k}$-space: $\hat{\cal N} \langle n'_\beta \,u_i
\rangle_{\bm k} = - \langle n'_\beta({\bm k})
\,u_i(-{\bm k}) \rangle / \tau(k)$ and $\hat{\cal
N} \langle \theta \,u_i \rangle_{\bm k} = -
\langle \theta({\bm k}) \,u_i(-{\bm k}) \rangle /
\tau(k)$. Taking into account that the
characteristic time scale of variation of the
second moments is much larger than the turbulent
time $\tau(k)$, we arrive at the following
steady-state solutions for the second moments:
\begin{eqnarray}
&& \langle n'_\beta({\bm k}) \,u_i(-{\bm k})
\rangle = -\tau(k) \, \Big[\nu_\beta \overline{W}
\langle C'({\bm k}) \,u_i(-{\bm k}) \rangle
\nonumber\\
&&\quad +\langle u_i({\bm k}) \, u_j(-{\bm k})
\rangle \, \Big(\nabla_j \overline{N}_\beta +
\overline{N}_\beta \, \nabla_j \ln
\overline{T}\Big) \Big],
\nonumber\\
 \label{E9}\\
&& \langle \theta({\bm k}) \,u_i(-{\bm k})
\rangle = -\tau(k) \, \Big[-q \overline{W}
\langle C'({\bm k}) \,u_i(-{\bm k}) \rangle
\nonumber\\
&&\quad +\gamma \langle u_i({\bm k}) \, u_j(-{\bm
k}) \rangle \, \nabla_j \overline{T} \Big],
 \label{E10}
\end{eqnarray}
where the flux $\langle C'({\bm k}) \,u_i(-{\bm
k}) \rangle$ is determined by Eq.~(\ref{D2}).
After integration in ${\bm k}$-space, we finally
arrive at the following equation for the
turbulent flux of reacting admixtures, $\langle
n'_\beta \, {\bm u} \rangle$, and the turbulent
heat flux, $\langle \theta \, {\bm u} \rangle$:
\begin{eqnarray}
\langle n'_\beta \, {\bm u} \rangle &=& -
D^T_\beta \, {\bm \nabla} \overline{N}_\beta +
\sum_{\lambda=1; \lambda\not=\beta}^{m} \, D^{\rm
MTD}_{\lambda}(\beta) \, {\bm \nabla}
\overline{N}_\lambda
\nonumber\\
&&\quad  + {\bm V}_{\rm eff} \,
\overline{N}_\beta ,
 \label{E11}\\
\langle \theta \, {\bm u} \rangle &=& - D^T \,
{\bm \nabla} \overline{T} -\sum_{\lambda=1}^{m}
D^{\rm TDE}_{\lambda} \, {\bm \nabla}
\overline{N}_\lambda .
 \label{E12}
\end{eqnarray}
In Eqs.~(\ref{E11}) and~(\ref{E12}) we used the
following notations:

\noindent $D^T_\beta$ is the coefficient of
turbulent diffusion of the number density of
admixtures,
\begin{eqnarray}
D^T_\beta=D^T_0 \, \left(1 - {\nu_\beta^2 \over
\overline{N}_\beta \, (\alpha_n-\alpha_{_{T}})}
 \, \Phi({\rm Da}_{_{\rm T}}) \right),
 \label{E14}
\end{eqnarray}
$\Phi({\rm Da}_{_{\rm T}})$ is the
non-dimensional function,
\begin{eqnarray}
\Phi({\rm Da}_{_{\rm T}})=1 - {1 \over {\rm
Da}_{_{\rm T}}} \, \left(1- {\ln(1 + 2{\rm
Da}_{_{\rm T}}) \over 2{\rm Da}_{_{\rm T}}}
\right);
 \label{E15}
\end{eqnarray}
$D^{\rm MTD}_{\lambda}(\beta)$ is the coefficient
of the mutual turbulent diffusion of the number
density of admixtures:
\begin{eqnarray}
D^{\rm MTD}_{\lambda}(\beta)= D^T_0 \, {\nu_\beta
\nu_\lambda \over \overline{N}_\lambda \,
(\alpha_n-\alpha_{_{T}})} \, \Phi({\rm Da}_{_{\rm
T}});
 \label{E16}
\end{eqnarray}
${\bm V}_{\rm eff}$ is the effective velocity of
the number density of admixtures due to the
turbulent thermal diffusion:
\begin{eqnarray}
{\bm V}_{\rm eff} = - D^{\rm TTD}_{\beta} \, {\bm
\nabla} \ln \overline{T},
 \label{E17}
\end{eqnarray}
$D^{\rm TTD}_{\beta}$ is the coefficient of
turbulent thermal diffusion,
\begin{eqnarray}
D^{\rm TTD}_{\beta} &=& D^T_0 \, \Big[1 -
{\nu_\beta \over \overline{N}_\beta \,
(\alpha_n-\alpha_{_{T}})} \, \Phi({\rm Da}_{_{\rm
T}})
\nonumber\\
&& \times \Big(\sum_{\lambda=1}^{m} \nu_\lambda +
\gamma {E_a\over R \overline{T}} \Big) \Big];
 \label{E18}
\end{eqnarray}
$D^{T}$ is the coefficient of turbulent diffusion
of the temperature,
\begin{eqnarray}
D^{T} &=& D^T_0 \, \gamma \, \Big[1 + {q \over
\gamma \, \overline{T} \,
(\alpha_n-\alpha_{_{T}})} \, \Phi({\rm Da}_{_{\rm
T}})
\nonumber\\
&& \times \Big(\sum_{\beta=1}^{m} \nu_\beta +
\gamma {E_a\over R \overline{T}} \Big) \Big];
 \label{E20}
\end{eqnarray}
$D^{\rm TDE}_{\lambda}$ is the coefficient that
describes the turbulent Duffor effect,
\begin{eqnarray}
D^{\rm TDE}_{\lambda} &=& D^T_0 \, {q \,
\nu_\lambda \over \overline{N}_\lambda \,
(\alpha_n-\alpha_{_{T}})} \, \Phi({\rm Da}_{_{\rm
 T}}) .
 \label{E21}
\end{eqnarray}
Here we used that $\int \tau_{\rm eff}(k) \tau(k)
\langle u_i({\bm k}) \, u_j(-{\bm k}) \rangle \,
d{\bm k} = D^T_0 \tau_c \, \Phi({\rm Da}_{_{\rm
T}}) \, \delta_{ij}$. The asymptotic behaviour of
the function $\Phi({\rm Da}_{_{\rm T}})$ is as
follows: when ${\rm Da}_{_{\rm T}} \ll 1$, the
function $\Phi({\rm Da}_{_{\rm T}})$ is
\begin{eqnarray}
\Phi({\rm Da}_{_{\rm T}}) = {4{\rm Da}_{_{\rm T}}
\over 3} + O\left({\rm Da}_{_{\rm T}}^2 \right),
 \label{E22}
\end{eqnarray}
while for ${\rm Da}_{_{\rm T}} \gg 1$, it is
\begin{eqnarray}
\Phi({\rm Da}_{_{\rm T}}) =1- {1 \over {\rm
Da}_{_{\rm T}}} \, \left(1- {\ln{\rm Da}_{_{\rm
T}}\over 2{\rm Da}_{_{\rm T}}} \right) .
 \label{E23}
\end{eqnarray}

Let us discuss the mechanisms of the effects that
are described by the different terms in
Eqs.~(\ref{E11}) and~(\ref{E12}) for the
turbulent flux of reacting admixtures, $\langle
n'_\beta \, {\bm u} \rangle$, and the turbulent
heat flux, $\langle \theta \, {\bm u} \rangle$.
In addition to the turbulent diffusion terms in
the turbulent flux of reacting admixtures, $-
D^T_\beta \, {\bm \nabla} \overline{N}_\beta$,
and the turbulent heat flux, $- D^T \, {\bm
\nabla} \overline{T}$, there are different
turbulent cross-effects that are discussed below.

The term ${\bm V}_{\rm eff} \, \overline{N}_\beta
= - \overline{N}_\beta \, D^{\rm TTD}_{\beta} \,
{\bm \nabla} \ln \overline{T}$ in the
expression~(\ref{E11}) for the turbulent flux of
reacting admixtures, $\langle n'_\beta \, {\bm u}
\rangle$ describes the phenomenon of turbulent
thermal diffusion. This effect has been predicted
theoretically \cite{EKR96,EKR00} and detected in
different laboratory experiments in stably and
unstably temperature-stratified turbulence
produced by oscillating grids or a multi-fan
generator \cite{BEE04,EEKR06,EKR10}. Turbulent
thermal diffusion has been also detected in
direct numerical simulations \cite{HKRB12} and is
shown to be important for atmospheric turbulence
with temperature inversions \cite{SSEKR09} and
for small-scale particle clustering in
temperature-stratified turbulence
\cite{EKR10,EKR13}.

The phenomenon of turbulent thermal diffusion in
temperature-stratified turbulence causes a
non-diffusive turbulent flux (i.e.,
non-counter-gradient transport) of gaseous
admixtures in the direction of the turbulent heat
flux and results in the formation of large-scale
inhomogeneities in the spatial distribution of
gaseous admixtures, so that admixtures are
accumulated in the vicinity of the mean
temperature minimum. A competition between
turbulent thermal diffusion and turbulent
diffusion determines the conditions for the
formation of large-scale gaseous clouds with the
characteristic scale that is much larger than the
integral scale of the turbulence and the
characteristic life-time that is much larger than
the characteristic turbulent time.

The physics of the accumulation of gaseous
admixtures in the vicinity of the maximum of the
mean fluid density (or the minimum of the mean
fluid temperature) can be explained as follows.
Let us assume that the fluid mean density
$\rho_2$ at point $2$ is larger than the fluid
mean density $\rho_1$ at point $1$. Consider two
small control volumes ``a'' and ``b'' located
between these two points, and let the direction
of the local turbulent velocity in volume ``a''
at some instant be the same as the direction of
the mean fluid density gradient $\bec\nabla \,
\rho$ (i.e., towards point $2$). Let the local
turbulent velocity in volume ``b'' at this
instant be directed opposite to the mean fluid
density gradient (i.e., towards point $1$).

In a low-Mach-number fluid flow with an imposed
mean temperature gradient (i.e., an imposed mean
fluid density gradient), one of the sources of
gaseous admixtures fluctuations, $n'_\beta
\propto - \tau_0 \, \overline{N}_\beta \,
(\bec\nabla {\bf \cdot} \, {\bm u})$, is caused
by a non-zero $\bec\nabla\cdot{\bm u} \approx -
({\bm u} \cdot \bec\nabla) \ln \rho \not=0$ [see
the second term on the right hand side of
Eq.~(\ref{EE5})]. Since fluctuations of the fluid
velocity ${\bm u}$ are positive in volume ``a''
and negative in volume ``b'', we have $\bec\nabla
{\bf \cdot} \, {\bm u} < 0$ in volume ``a'', and
$\bec\nabla {\bf \cdot} \, {\bm u} > 0$ in volume
``b''. Therefore, the fluctuations of the gaseous
admixtures number density $n'_\beta \propto -
\tau_0 \, \overline{N}_\beta \, (\bec\nabla {\bf
\cdot} \, {\bm u})$ are positive in volume ``a''
and negative in volume ``b''. However, the flux
of gaseous admixtures $n'_\beta \, {\bm u}$ is
positive in volume ``a'' (i.e., it is directed
toward point $2$), and it is also positive in
volume ``b'' (because both fluctuations of fluid
velocity and number density of particles are
negative in volume ``b''). Therefore, the mean
flux of gaseous admixtures $\langle n'_\beta \,
{\bm u} \rangle$ is directed, as is the mean
fluid density gradient $\bec\nabla \, \rho$,
toward point~2. This forms large-scale
heterogeneous structures of gaseous admixtures in
regions with a mean fluid density maximum.

The term $\sum_{\lambda=1; \lambda\not=\beta}^{m}
\, D^{\rm MTD}_{\lambda}(\beta) \, {\bm \nabla}
\overline{N}_\lambda$ in the
expression~(\ref{E11}) for the turbulent flux of
reacting admixtures, $\langle n'_\beta \, {\bm u}
\rangle$ describes the  mutual turbulent
diffusion of admixtures. Let us discuss the
mechanism of this effect. It is known in
irreversible thermodynamics that the mutual
molecular diffusion of admixtures is caused by
interaction between gaseous admixtures due to
collisions of molecules of the admixtures. In
turbulent flow with chemical reactions
inhomogeneities of the number density of one of
the reagents causes fast change (during the
chemical reaction time scale $\tau_{c})$ of the
number density of other components due to the
shift from the chemical equilibrium.

The inhomogeneities of the number density of the
admixture cause heat release (or absorption) due
to the thermal effects of the chemical reactions,
i.e., additional non-diffusive turbulent heat
flux, that is determined by the term
$-\sum_{\lambda=1}^{m} D^{\rm TDE}_{\lambda} \,
{\bm \nabla} \overline{N}_\lambda$ in the
expression~(\ref{E12}) for $\langle \theta \,
{\bm u} \rangle$. This flux can be interpreted as
turbulent analogue of the molecular Duffor effect
known in irreversible thermodynamics.

\begin{figure}
\vspace*{1mm} \centering
\includegraphics[width=8.5cm]{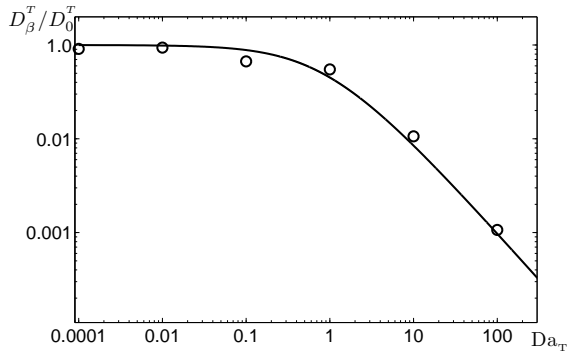}
\caption{\label{Fig1} Comparison of the theoretical dependence
of turbulent diffusion coefficient $D^T_\beta/ D_0^T$ versus
turbulent Damk\"{o}hler number ${\rm Da}_{_{\rm T}}$ with the corresponding
results of MFS performed in \cite{BH11}.}
\end{figure}

\section{Comparison with numerical simulations}

In this section we compare the obtained theoretical dependence
of turbulent diffusion coefficient $D^T_\beta/ D_0^T$ versus
turbulent Damk\"{o}hler number ${\rm Da}_{_{\rm T}}$ with the corresponding
results of MFS performed in \cite{BH11}, where
a reactive front propagation in a turbulent flow was studied
using the Kolmogorov-Petrovskii-Piskunov-Fisher
equation.
To describe the interaction with a turbulent velocity field
an advection term  was added to this equation,
so that advection-reaction-diffusion equation reads \cite{BH11}:
\begin{eqnarray}
\frac{\partial n}{\partial t} + {\bm \nabla
\cdot} (n \,{\bm v}) = {n \over \tau_c} \left(1- {n \over n_0} \right) + D \Delta n ,
 \label{L1}
\end{eqnarray}
where $n = n_0$ is a stable equilibrium solution of Eq.~(\ref{L1}).
After averaging Eq.~(\ref{L1}) the following mean-field equation
was obtained in \cite{BH11}:
\begin{eqnarray}
\frac{\partial \overline{N}}{\partial t} + \tau \frac{\partial^2 \overline{N}}{\partial t^2}
= {\overline{N} \over \tau_c} \left(1- {\overline{N} \over n_0} \right)
+ D_{\rm T} \Delta \overline{N} ,
 \label{L2}
\end{eqnarray}
where $D_{\rm T}$ is the sum of turbulent and molecular diffusion coefficients.
The second term in the left hand side of Eq.~(\ref{L2}) that is proportional
to the memory time $\tau$, determines the memory effects of turbulent diffusion.
One-dimensional Eq.~(\ref{L2}) was solved numerically in \cite{BH11}
to determine the dependence of the front speed, $s_{_{\rm T}}$ on turbulent Damk\"{o}hler
numbers. Here the reaction speed, $s_{_{\rm T}} =(d/dt) \int (\overline{N} / n_0) \,dz$,
is determined by differentiating the concentration integrated
over the whole domain, and approximating the asymptotic front speed with the
value at the time when the front has reached the other
end of the computational domain.

Comparison of the theoretical dependence
of turbulent diffusion coefficient versus
turbulent Damk\"{o}hler number with the corresponding
results of MFS is shown in Fig.~\ref{Fig1}, and it demonstrate a very good agreement
between the theoretical predictions [see Eqs.~(\ref{E14}) and~(\ref{E15}) with $\nu_\beta=1$
and $\alpha_n=1/\overline{N}_\beta \gg \alpha_{_{T}}$] and numerical simulations of \cite{BH11}.
To obtain the function $D^T_\beta({\rm Da}_{_{\rm T}})$ we have taken into account
that $s_{_{\rm T}} =2 (D^T_\beta/\tau_c)^{1/2}$ (see \cite{EKR98,BH11}).
Note that a detailed comparison of the theoretical results with DNS requires
a specially designed DNS that is a subject of separate ongoing
study.

\section{Discussion and Conclusions}

In this study we investigated effects of the
chemical reactions on turbulent transport and
turbulent diffusion of gaseous admixtures. To
elucidate physics of the obtained results we
consider examples of chemical reactions
proceeding in a stoichiometric mixture. For a
small concentration of reactive admixtures,
$\overline{N}_\beta \ll \overline{N}_f$, the
characteristic chemical time $\tau_c$ varies from
$10^{-3}$ s to $10^{-2}$ s, where
$\overline{N}_f$ is the ambient fluid number
density. For typical values of turbulent velocity
in atmospheric flows $u_0 = 1 $ m/s and integral
scale $\ell = 100$ m, we obtain characteristic
turbulent time $\tau_0 = \ell /u_0 = 10^2$ s, so
that the case of large turbulent Damk\"{o}hler
numbers, ${\rm Da}_{_{\rm T}} = \tau_0/\tau_c \gg
1$ is of the main physical interest. Let us also
estimate the ratio $\alpha_n/\alpha_{_{T}}$ in
the expression~(\ref{E14}) for the coefficients
of turbulent diffusion. Using Eq.~(\ref{B9}), we
rewrite equation for $\alpha_{_{T}}$ in the
following form:
\begin{eqnarray}
\alpha_{_{T}} \approx {T_f \over \overline{N}_f}
\, \left({E_a\over R\overline{T}^2}\right) .
 \label{F1}
\end{eqnarray}
We take into account that the temperature of the
reaction products equals the fluid temperature,
$\overline{T} = T_f$, the reactive species are
strongly diluted, $\overline{N}_\beta \ll
\overline{N}_f$, and $E_a / R\overline {T}$
varies in the range from 10 to 100. For a small
enough concentration of the reactive admixtures,
$\alpha_n/\alpha_{_{T}} \sim (E_a/
R\overline{T})^{-1}
(\overline{N}_f/\overline{N}_\beta) \gg 1$.

The stoichiometric coefficient $\nu_\beta$  in
Eq.~(\ref{B1}) is known as the order of the
reaction with respect to species $\beta$. In
practice the overall order of the reaction is
defined as the sum of the exponents of the
concentrations in the reaction rate,
$\nu_c=\sum_{\beta=1}^m \nu_\beta$. For a
simplified model of a single-step reaction the
overall order of the reaction is the molecularity
of the reaction, indicating the number of
particles entering the reaction. In general the
overall order of most chemical reactions is 2 or
3, though for complex reactions the overall order
of the reaction can be fractional one
\cite{SP06,ML08}.

Let us consider first the simplest chemical
reaction $A \to B$, assuming a large turbulent
Damk\"{o}hler numbers, ${\rm Da}_{_{\rm T}} \gg
1$. An example of such chemical reaction is the
dissociation: $O_2 \to O + O$. As follows from
Eqs.~(\ref{E14}) and~(\ref{E23}), turbulent
diffusion of the number density of admixtures is
$D^T_\beta = D_0^T/{\rm Da}_{_{\rm T}} = \tau_{c}
u_0^2/3$, which means that the turbulent
diffusion of admixture is determined by the
chemical time.  This is in agreement with the
result obtained in \cite{EKR98} where
path-integral approach in a turbulence model with
a very short correlation time was used. The
underlying physics of this phenomena is quite
transparent. For a simple first-order chemical
reaction $A \to B$ the species $A$ of the
reactive admixture are consumed and their
concentration decreases much faster during the
chemical reaction, so that the usual turbulent
diffusion based on the turbulent time $\tau_0 \gg
\tau_c$, does not contribute to the mass flux of
a reagent $A$. The turbulent diffusion during the
turnover time of the turbulent eddies is
effective only for the product of reaction, $B$.
Applicability of the obtained results requires
the condition ${\rm Pe} /{\rm Da}_{_{\rm T}} \gg
1$ to be satisfied, where ${\rm Pe}= u_0^2 \tau_0
/3$ is the Peclet number.

For multi-component second-order or third-order
chemical reactions, the impact of chemistry on
the turbulent diffusion is more complicated. Let
us determine the turbulent diffusion coefficients
for the second-order chemical reaction that is
determined by the following equation:
\begin{eqnarray}
A + B \to C + D .
 \label{E45}
\end{eqnarray}
An example of such chemical reaction is $H + O_2
\to OH + O$. The numbers of species in
Eq.~(\ref{E45}) are stoichiometric coefficients,
which define number of moles participating in the
reaction. The stoichiometric reaction whereby the
initial substances are taken in a proportion such
that the chemical transformation fully converts
them into the reaction products, can proceed as
the inverse reaction also.  For the reaction
given by Eq.~(\ref{E45}), we obtain $\alpha_n =
2/\overline{N}$, where $\overline{N}_A =
\overline{N}_B \equiv \overline{N}$. On the other
hand, using Eq.~(\ref{E14}) for the turbulent
diffusion coefficients of species $A$ and $B$, we
obtain
\begin{eqnarray}
D_{A,B}^T = \frac{1}{2}\, D_0^T \, \left(1 + {\rm
Da}_{_{\rm T}}^{-1}\right) .
 \label{E46}
\end{eqnarray}
Correspondingly for the third-order reaction, $A
+ B + C \to D + E$, we find
\begin{eqnarray}
D_{A,B,C}^T = \frac{2}{3} \, D_0^T \, \left(1 +
{\rm Da}_{_{\rm T}}^{-1}\right),
 \label{E47}
\end{eqnarray}
where we have taken into account that in this
case $\alpha_n = 3/\overline{N}$.

Consider the stoichiometric third-order reaction
with different stoichiometric coefficients of the
reagents $2A + B \to 2C$ (for example, the
chemical reaction  $2C + O_2 \to 2CO$ or $2 H_2 +
O_2 \to 2H_2O$). In this case $\alpha_n =
2/\overline{N}$, and the turbulent diffusion
coefficients of species $A$ and $B$ are as
follows:
\begin{eqnarray}
D_A^T = D_0^T/{\rm Da}_{_{\rm T}}, \quad D_B^T =
\frac{1}{2} \, D_0^T \, \left(1 + {\rm Da}_{_{\rm
T}}^{-1}\right) .
 \label{E48}
\end{eqnarray}
Since the species $A$ have a larger
stoichiometric coefficient and, correspondingly,
larger number of moles participating in the
chemical reaction, they are consumed more
effectively in the reaction and the turbulent
diffusion coefficient for the species $A$
decreases much stronger than that for species
$B$.

Note that the derived in the present study
theoretical dependence of turbulent diffusion
coefficient versus the turbulent Damk\"{o}hler number
is in a good agreement with that obtained in \cite{BH11} using MFS
of a reactive front propagating in a turbulent flow and
described by the Kolmogorov-Petrovskii-Piskunov-Fisher equation.

The turbulent thermal diffusion coefficient in
the case of a large turbulent Damk\"{o}hler
numbers, ${\rm Da}_{_{\rm T}} \gg 1$, is
\begin{eqnarray}
D^{\rm TTD}_{\beta} = D_0^T \, \left[1 -
\nu_\beta \, {\overline{N}_\beta \over
\overline{N}_f} \,\left({E_a \over R\overline{T}}
\right) \, \left[1 - O({\rm Da}_{_{\rm
T}}^{-1})\right] \right],
 \nonumber\\
 \label{E49}
\end{eqnarray}
which implies that the turbulent thermal
diffusion is only slightly sensitive to the
chemical reaction energy release in the case of
small concentration of the reactive admixture,
$(\overline{N}_\beta/\overline{N}_f)({E_a}/R\overline{T})
\ll 1$.

The mutual turbulent diffusion of the number
density of admixtures (determined by
$D_\beta^{\rm MTD})$  and the turbulent Duffor
effect (determined by $D_\beta^{\rm TDE})$ are
caused only by the chemical reaction, see
Eqs.~(\ref{E16}) and~(\ref{E21}). The mechanisms
of these cross-effects (e.g., heat flux caused by
concentration gradient, Dufour-effect; or mutual
diffusion of the number density of gaseous
admixtures; or turbulent thermal diffusion) are
different from molecular cross effects.

The effect of strong suppression of turbulent
diffusion also holds for the heterogeneous phase
transitions. It is plausible that this effect
explains the existence a sharp boundary of the
clouds containing very small droplets (of the
order of several microns), and a visible
diffusive boundary for raindrop clouds consisting
of 300 - 500 microns droplets.

It should be noticed that in the case of
non-stoichiometric reaction, when the substances
with higher stoichiometry (molecularity) [$A$ in
Eq.~(\ref{E48})] is excessive in the initial
mixture, e.g., appears in an amount larger than
that required according to the stoichiometric
equation, the turbulent diffusion coefficient of
the species $A$ tends to zero, so molecular
diffusion can be important.

\begin{acknowledgements}
This research was supported in part by the Israel
Science Foundation governed by the Israeli
Academy of Sciences (Grant 259/07, TE, NK, IR),
the Russian Government Mega Grant (Grant
11.G34.31.0048, NK, IR), the Research Council
of Norway under the FRINATEK (Grant 231444, NK,
ML, IR), by the Grant of Russian Ministry of
Science and Education (ML), and by Ben-Gurion
University Fellowship for senior visiting
scientists (ML).
\end{acknowledgements}

\end{document}